\documentclass{aa}

\usepackage{array}
\usepackage{graphics}
\usepackage{times}

\begin{document}


\thesaurus{10             
          (02.14.1        
           09.01.1,       
           10.01.1,       
           10.08.1)}      

\title{Metal-poor halo stars as tracers of ISM mixing processes during
       halo formation}

\titlerunning{Metal-poor halo stars as tracers of ISM mixing processes}

\author{D.~Argast \inst{1, 2}, M.~Samland \inst{1}, O.~E.~Gerhard \inst{1}
        \and F.-K.~Thielemann \inst{2}}

\authorrunning{D.~Argast et al.}

\institute{Astronomisches Institut der Universit\"at Basel,
           Venusstrasse 7, CH-4102 Binningen, Switzerland (argast@astro.unibas.ch)
\and       Institut f\"ur Physik der Universit\"at Basel,
           Klingelbergstrasse 82, CH-4056 Basel, Switzerland
           (fkt@quasar.physik.unibas.ch)}

\date{Received \dots, accepted \dots}

\maketitle


\begin{abstract}
We introduce a stochastic halo formation model to compute the early chemical
enrichment of the interstellar medium (ISM) of the halo. Contrary to 1-zone
chemical evolution models, we are able to resolve local inhomogeneities in the ISM
caused by single core-collapse supernovae. These inhomogeneities lead to different
element abundance patterns in very metal-poor stars, which can be seen as scatter
in the abundances of halo stars with metallicities [Fe/H] $< 2.0$.

The early chemical evolution of the halo proceeds in different enrichment phases:
At [Fe/H] $< -3.0$, the halo ISM is unmixed and dominated by local inhomogeneities
caused by individual core-collapse supernova (SN) events. For metallicities [Fe/H]
$>-2.0$ the halo ISM is well mixed, showing an element abundance pattern
integrated over the initial mass function. In the range $-3.0 <$ [Fe/H] $< -2.0$ a
continuous transition from the unmixed to the well mixed ISM occurs.

For some elements (Si, Ca, Eu), the scatter in the element-to-iron ratio [El/Fe]
of metal-poor halo stars can be reproduced. Stellar yields of other elements
predict a scatter which, compared to the observations, is too large (O, Mg) or too
small (Ni). Cr and Mn show a decreasing trend for lower metallicities, which can
not be explained by metallicity independent yields, provided that the mixing of
the ejecta with the interstellar medium does not depend on progenitor mass. This
demonstrates the need for revised, self-consistent SN yields.

Finally, we discuss the metallicity distribution in the model. Compared to the 28
very metal-poor stars observed with metallicities in the range $-4.0 <$ [Fe/H] $<
-3.0$, no star is known with confirmed metallicity [Fe/H] $< -4.0$, while our
model predicts $5\pm2$ stars with [Fe/H] $< -4.0$. These should be present if the
halo ISM started at primordial metallicities and no pre-enrichment by population
III stars occurred.

\end{abstract}


\section{Introduction}

The low metal abundances and high peculiar velocities of halo stars indicate that
the halo is an old, if not the oldest component of the Milky Way. Age
determinations of globular clusters and halo field stars point to an age of 14-15
billion years with no detectable age gradient with galactocentric distance (Harris
et al. \cite{ha97}). The halo has therefore special significance for the formation
of the Milky Way. There are two quantities which play important r\^oles in the
investigation of the formation of the halo: the orbits of halo stars and their
chemical composition. Since halo stars form a collisionless system, their orbits
contain information about the dynamics at the time of star formation and thus the
formation of the halo (e.g. Carney et al. \cite{ca96}; Chiba \& Yoshii
\cite{ch98}). Information about the chemical composition of the interstellar
medium (ISM) of the halo is more direct. The element abundances observed in low
mass halo stars directly reflect the chemical abundances and the chemical
inhomogeneity of the ISM during halo formation (McWilliam \cite{mw97}).

Examinations of element abundance ratios as function of metallicity [Fe/H] show
that star-to-star differences rise with decreasing metallicity (Ryan et al.
\cite{ry96}). Most of the chemical elements are ejected during supernovae Type~II
(SNe~II) explosions. The enrichment of the halo depends on how many SNe~II explode
and how effectively the ejected gas is mixed with the surrounding ISM. If the
ejected metals are distributed over a large volume, a spatially homogeneous
enrichment takes place. If the mixing volume is small, the ISM in the vicinity of
a core-collapse supernova (SN~II) is highly enriched, while large parts of the
halo gas remain metal-poor. In this case the ISM is chemically highly
inhomogeneous and newly formed stars are of different chemical composition,
depending on where they form. In this scenario one should moreover expect that the
metal-poorest stars have chemical compositions corresponding to the stellar yields
of single SNe~II (Ryan et al. \cite{ry96}).

In this paper we present a stochastic chemical evolution model and investigate the
inhomogeneous enrichment of the halo ISM. The description of the model is given in
Sect.~\ref{model}, followed by an overview of theoretical SN~II yields and their
uncertainties in Sect.~\ref{nucleo}. The employed observational data is presented
in Sect.~\ref{obsdat}. The results of our model and the conclusions are given in
Sect.~\ref{results} and \ref{conclusions}, respectively.


\section{The Model}
\label{model}

Very metal-poor halo stars show a great diversity in their element abundances and
therefore a scatter in their element-to-iron ratios [El/Fe] of order 1 dex. This
scatter gradually decreases at higher metallicities until a mean element abundance
is reached which corresponds to the [El/Fe] ratio of the stellar yields integrated
over the initial mass function (IMF). The aim of our stochastic halo formation
model is to understand the trends seen in the observations and to investigate how
the metal-poor interstellar medium (ISM) in the halo evolves chemically.

Our fully 3D-code, contrary to 1-zone chemical evolution models, enables us to
resolve local inhomogeneities in the ISM with a spatial resolution of 50 pc. All
in all, we model a volume of (2.5 kpc)$^3$, divided into $50^3$ cells. Every cell
of our grid contains detailed information about the enclosed ISM and the mass
distribution of stars. We consider simultaneously the evolution of nine elements,
the $\alpha$-elements O, Mg, Si and Ca, the iron-peak elements Cr, Mn, Fe and Ni
and the r-process element Eu.

Our initial conditions assume a halo ISM consisting of a homogeneously distributed
single gas phase with primordial abundances and a density of 0.25 particles per
cm$^3$, which gives a total mass of about $10^8 \, \mathrm{M}_{\sun}$ in a volume
of (2.5 kpc)$^3$. We adopt a constant time-step of $10^6$ years since it has to be
longer than the dynamical evolution of a supernova (SN) remnant and shorter than
the lifetime of the most massive stars. At each time-step 20\,000 cells are chosen
randomly and independently of each other and of the state of the enclosed
ISM. Each selected cell may create a star with a probability proportional to the
square of the local ISM density (Larson \cite{la88}). The number of stars formed
per time-step is the product of the number of cells tested with the probability of
star formation in each cell. Various combinations of these parameters are possible
to achieve a given SFR; the choice of 20\,000 cells proved computationally
convenient. The absolute value of the SFR influences the \emph{time-scale} of the
enrichment process (cf. Sect.~\ref{mixing}), but \emph{not} the evolution of
[El/Fe]-ratios as function of [Fe/H]. Therefore, the main results of this paper
are insensitive to the values of these parameters.

The mass of a newly formed star is chosen randomly from a Salpeter IMF. The lower
and upper mass limits of the IMF are taken to be $0.1 \, \mathrm{M}_{\sun}$ and
$50 \, \mathrm{M}_{\sun}$, respectively. About 5000 stars are formed on average
during each step. Newly born stars inherit the abundance pattern of the ISM out of
which they form, carrying therefore information about the state of the ISM at the
place and time of their birth. To determine the lifetime of a star an
approximation to the metallicity dependent mass-lifetime relation of the
\emph{Geneva Stellar Evolution and Nucleosynthesis Group} (cf. Schaller et
al. \cite{sl92}; Schaerer et al. \cite{sr93a}; Schaerer et al. \cite{sr93b};
Charbonnel et al. \cite{cb93}) is used, given by
\begin{eqnarray*}
\log \left( T \right) &=& \left( 3.79 + 0.24 \cdot Z \right) - \left( 3.10 +
0.35 \cdot Z \right) \, \log \left( M \right) \\
&&+ \left( 0.74 + 0.11 \cdot Z \right) \, \log^2 \left( M \right),
\end{eqnarray*}
where $T$ is the lifetime in units of $10^6$ yr, $Z$ the metallicity in units of
solar metallicity $Z_{\sun}$ and $M$ the mass in units of solar masses
$\mathrm{M}_{\sun}$.

Stars in a range of $10-50 \, \mathrm{M}_{\sun}$ will explode as core-collapse
supernovae (SNe~II), resulting in an enrichment of the neighbouring ISM. Stellar
yields are taken from Thielemann et al. (\cite{th96}) and Nomoto et al.
(\cite{no97}) for all elements except Eu. Since there are no theoretical
predictions of stellar Eu yields, we use the indirectly deduced yields of
Tsujimoto \& Shigeyama (\cite{ts98}) which assume that r-process elements
originate from SNe~II (see the discussion of stellar yields in
Sect.~\ref{nucleo}). We linearly interpolate the stellar yields given in these
papers, since we use a finer mass-grid in our simulation. For SNe with masses
below $13 \, \mathrm{M}_{\sun}$ stellar yields are not available. Since the
nucleosynthesis models show declining yields towards low progenitor masses, we
have for the interpolation arbitrarily set the yields of a $10 \,
\mathrm{M}_{\sun}$ SN to one thousandth of those of a $13 \, \mathrm{M}_{\sun}$
SN. The interpolation gives IMF averaged values of the [El/Fe] ratios, which are
in good agreement with the observed mean values of metal-poor stars in all
elements except Ca, which shows a [Ca/Fe] ratio that is about 0.3 dex lower than
the observed mean. We do not include supernovae (SNe) of Type~Ia, since we are
only interested in the very early enrichment of the halo ISM, which is dominated
by SNe of Type~II.

Intermediate mass stars will evolve to planetary nebulae, returning only slightly
enriched material in the course of their evolution. This locally influences the
enrichment pattern of the gas, since metal-poor material is returned into the
evolved and enriched ISM. It will not change the element abundances [El/H]
significantly, but can affect the local element-to-iron ratios [El/Fe]
considerably. Low mass stars do not evolve significantly during the considered
time. In our model, they serve to lock up part of the gas mass, affecting
therefore the local element abundances [El/H] in the ISM.

Since the explosion energy of a core-collapse supernova (SN~II) depends only
slightly on the mass of its progenitor (Woosley \& Weaver \cite{ww95}; Thielemann
et al. \cite{th96}), every SN~II sweeps up a constant mass of about $5 \times 10^4
\, \mathrm{M}_{\sun}$ of gas (Ryan et al. \cite{ry96}; Shigeyama \& Tsujimoto
\cite{sh98}). In our model the radius of the SN remnant then is computed from the
local density of the ISM and lies typically between 100~pc and 200~pc. The ejecta
of the SN~II and all the swept up, enriched material are condensed in a spherical
shell which is assumed to be chemically well mixed. The material in the shell
subsequently mixes with the ISM of the cells where the expansion of the remnant
stopped. The interior of the remnant, where all the material was swept up, is
assumed to be filled with about $5 \, \mathrm{M}_{\sun}$ of dilute gas from the SN
event with the corresponding metal abundances. This gas is unable to form stars
until it is swept up by another SN event and mixed with the surrounding ISM. Thus
this material contributes to the enrichment only after some delay.

The star formation rate of cells influenced by the remnant will rise, since their
density is higher than the average density of a cell and the probability to form a
star is assumed to be proportional to the square of the local density. It is still
possible to form stars in the field, but cells that are influenced by a SN remnant
are favoured. Stars which form out of material enriched by a single SN inherit its
abundances and therefore show an abundance pattern which is characteristic for
this particular progenitor mass. The most metal-poor stars that form out of
material which was enriched by only one SN would therefore allow to reconstruct
the stellar yields of single core-collapse SN, if the average swept up mass and
the absolute yield of one element were known (Shigeyama \& Tsujimoto \cite{sh98}).

The SN remnant expansion is the only dynamical process taken into account in our
model. Therefore, this model has the least possible mixing of the ISM. This is the
opposite limit as in the case of closed box models, which assume a complete mixing
of the ISM at all times and are therefore not able to explain the scatter seen at
low metallicities. We continue our calculation up to an averaged iron abundance of
[Fe/H] $= -1.0$. At this metallicity SN events of Type~Ia, which we have not
included in our model, would start to influence the ISM.


\section{SN II yields and their uncertainties}
\label{nucleo}

In the present investigation we make use of the nucleosynthesis results by
Thielemann et al. (\cite{th96}) and Nomoto et al. (\cite{no97}). Here we want to
give a short summary of the key features together with an assessment of the
uncertainties by comparing with available independent calculations. The
synthesized elements form three different classes which are sensitive to different
aspects of the stellar models and supernovae explosion mechanism: (1) stellar
evolution, (2) stellar evolution plus the explosion energy, and (3) details of the
explosion mechanism which includes aspects of stellar evolution determining the
size of the collapsing Fe-core. Due to reaction equilibria obtained in explosive
burning, the results do not show a strong sensitivity to the applied reaction rate
library (Hoffman et al. \cite{ho99}).

1: The abundances of C, O, Ne, and Mg originate from the unaltered (essentially
only hydrostatically processed) C-core and from explosive Ne/C-burning. They are
mainly dependent on the structure and zone sizes of the pre-explosion models
resulting from stellar evolution. These zones and therefore the amount of ejected
mass varies strongly over the progenitor mass range. O, Ne, and Mg vary by a
factor of 10-20 between a $13 \, \mathrm{M}_{\sun}$ and a $25 \,
\mathrm{M}_{\sun}$ progenitor star. This behaviour can vary with the treatment of
stellar evolution and is strongly related to the amount and method of mixing in
unstable layers. Woosley \& Weaver (\cite{ww95}) employ the Ledoux criterion with
semiconvection for Schwarzschild-unstable but Ledoux-stable layers. Nomoto \&
Hashimoto (\cite{no88}) make use of the Schwarzschild criterion for convection
(neglecting composition gradients) which ensures mixing over more extended regions
than the Ledoux criterion. The Schwarzschild criterion causes larger convective
cores (see also Chieffi et al. \cite{ch98}) which leads to larger $^{16}$O,
$^{20}$Ne, and $^{24}$Mg yields, the latter being also dependent on the
$^{12}$C($\alpha,\gamma)$ rate (Langer \& Henkel \cite{la95}). In addition, it is
important to know the mixing velocity in unstable regions. Recent calculations by
Umeda et al. (\cite{um99}), within the diffusion approximation for mixing (Spruit
\cite{sp92}; Saio \& Nomoto \cite{sa98}) but with a remaining free parameter -
permitted to vary between 0 and 1 - show that the Woosley \& Weaver (\cite{ww95})
results can be reproduced with a small choice of this parameter of 0.05. A further
effect is due to rotation. When also treating rotation correctly (Langer et
al. \cite{la97}; Talon et al. \cite{ta97}; Meynet \& Maeder \cite{me97}; Heger et
al. \cite{he99}), rotational instabilities lead to additional mixing which can
bring the models making use of the Ledoux criterion closer to those evolved with
the Schwarzschild criterion and the compositions closer to those obtained with
instantaneous mixing (high mixing velocities). Thus, the amount of mixing (being
influenced by the mixing criterion utilized, the mixing velocity, and rotation)
determines in stellar evolution the size of the C/O core. While the yield of O can
be fixed with a combination of the still uncertain $^{12}$C($\alpha,\gamma)$ rate
(Buchmann \cite{bu96}, \cite{bu97}) and a mixing description, the yields of Ne and
Mg depend on the extent of mixing. Recent galactic chemical (but not dynamic)
evolution calculations (Thomas et al. \cite{th98}; Matteucci et al. \cite{ma99};
Chiappini et al. \cite{ch99}) prefer apparently a larger extent of mixing (caused
by either of the effects mentioned above) in order to reproduce the observed Mg in
low metallicity stars.

2: The amount of mass for the elements Si, S, Ar and Ca, originating from
explosive O- and Si-burning, is almost the same for all massive stars in the
Thielemann et al. (\cite{th96}) models. They do not show the strong progenitor
mass dependence of C, O, Ne, and Mg. Si has some contribution from hydrostatic
burning and varies by a factor of 2-3. Thus, the first set of elements (C, O, Ne,
Mg) tests the stellar progenitor models, while the second set (Si, S, Ar, Ca)
tests the progenitor models and the explosion energy, because the amount of
explosive burning depends on the structure of the model plus the energy of the
shock wave which passes through it. Present models make use of an artificially
induced shock wave via thermal energy deposition (Thielemann et al. \cite{th96})
or a piston (Woosley \& Weaver \cite{ww95}) with shock energies which lead, after
the reduction of the gravitational binding of ejected matter, to a given kinetic
energy. In our models this is an average energy of $10^{51}$ erg, known from
remnant observations, which does not reflect possible explosion energy variations
as a function of progenitor mass. The apparent underproduction of Ca seen in some
chemical evolution calculations (e.g. Thomas et al. \cite{th98}; Matteucci et
al. \cite{ma99}; Chiappini et al. \cite{ch99}) could apparently be solved by a
progenitor mass dependent explosion mechanism and energy.

3: The amount of Fe-group nuclei ejected (which includes also one of the so-called
alpha elements, i.e. Ti) and their relative composition depends directly on the
explosion mechanism, connected also to the size of the collapsing Fe-core.
Observational checks of individual supernovae are presently still required to test
the detailed working of a supernova. The present situation is still uncertain and
depends on Fe-cores from stellar evolution, the supranuclear equation of state and
maximum neutron star mass, related to the total amount of gravitational binding
energy release of the collapsed protoneutron star, the resulting total amount of
neutrinos, and the time release (luminosity), dependent on neutrino transport via
numerical treatment, convective transport, and opacities (Burrows \cite{bu90};
Herant et al. \cite{he94}; Janka \& M\"uller \cite{ja95}, \cite{ja96}; Keil \&
Janka \cite{ke95}; Burrows et al. \cite{bu95}, \cite{bu96}; Reddy \& Prakash
\cite{re97}; Burrows \& Sawyer \cite{bu98}; Mezzacappa et al. \cite{mezz98};
Messer et al. \cite{mess98}; Yamada et al. \cite{ya99}; Pons et al. \cite{po99}).
Three types of uncertainties are inherent in the Fe-group ejecta, related to (i)
the total amount of Fe (-group) nuclei ejected and the mass cut between neutron
star and ejecta, mostly measured by $^{56}$Ni decaying to $^{56}$Fe, (ii) the
total explosion energy which influences the entropy of the ejecta and with it the
degree of alpha-rich freeze-out from explosive Si-burning and the abundances of
radioactive $^{44}$Ti as well as $^{48}$Cr, the latter decaying later to $^{48}$Ti
and being responsible for elemental Ti, and (iii) finally the neutron richness or
$Y_{\mathrm{e}}$=$<Z/A>$ of the ejecta, dependent on stellar structure and the
delay time between collapse and explosion. $Y_{\mathrm{e}}$ influences strongly
the ratios of isotopes 57/56 in Ni (Co, Fe) and the overall elemental Ni/Fe
ratio. The latter being dominated by $^{58}$Ni and $^{56}$Fe. The position of the
mass cut has also a side effect (besides determining the total amount of
$^{56}$Ni/Fe), it influences the ratio of abundances from alpha-rich freeze-out
and incomplete Si-burning, affecting in this way the abundances of the elements Mn
($^{55}$Co decay), Cr ($^{52}$Fe decay) and Co ($^{59}$Cu decay) as discussed in
Nakamura et al. (\cite{na99}).

There is limited direct observational information from individual supernovae with
known progenitors (SN 1987A, SN 1993J, 1997D?, 1996N?, 1994I) and possible
hypernovae (SN1997ef, 1998bw), leading to direct O, Ti or Fe (Ni) observations
(e.g. Shigeyama \& Nomoto \cite{sh90}; Iwamoto et al. \cite{iw94}, \cite{iw98},
\cite{iw99a}, \cite{iw99b}; Turatto et al. \cite{tu98}; Sollerman et al.
\cite{so98}; Kozma \& Fransson \cite{ko98}; Bouchet et al. \cite{bo91}; Suntzeff
et al. \cite{su92}). As explosive nucleosynthesis calculations cannot presently
rely on self-consistent explosion models, the position of the mass cut is in all
cases an assumption and has mostly been normalized to observations of SN 1987A.
Whether there is a decline in Fe-ejecta as a function of progenitor mass (as
assumed in Thielemann et al. \cite{th96}) or actually an increase (Woosley \&
Weaver \cite{ww95}) or a more complex rise, maximum and decline (Nakamura et
al. \cite{na99}) is not really understood. Thus, the results by Thielemann et
al. (\cite{th96}) utilized here are showing the correct IMF integrated behaviour
of e.g. Si/Fe, but one has to keep in mind that e.g. O/Fe, Mg/Fe, Si/Fe, Ca/Fe
yields of individual supernovae could be quite uncertain and even show an
incorrect progenitor mass dependence or a larger scatter than (yet unknown)
realistic models. Ratios within the Fe-group (like e.g. Ni/Fe) have been obtained
by mass cut positions which reproduce the solar ratios. Thus, the theoretical
yields might show already the average values and a much smaller scatter than some
observations (see e.g. Henry \cite{he84}). Later work attempted to choose mass
cuts in order to represent some specific element trends like e.g. in Cr/Fe, Co/Fe
or Mn/Fe (Nakamura et al. \cite{na99}).

In general we should keep in mind that as long as the explosion mechanism is not
completely and quantitatively understood yet, one has to assume a position of the
mass cut. Dependent on that position, which is a function of explosion energy and
the delay time between collapse and final explosion, the total amount of Fe-group
matter can vary strongly, Ti-yields can vary strongly due to the attained
explosion energy and entropy, and the ejected mass zones will have a variation in
neutron excess which automatically changes relative abundances within the
Fe-group, especially the Ni/Fe element ratio.

4: r-Process Yields. SNe II have long been expected to be the source of r-process
elements. Some recent calculations seemed to be able to reproduce the solar
r-process abundances well in the high entropy neutrino wind, emitted from the hot
protoneutron star after the SN II explosion (Takahashi et al. \cite{ta94}; Woosley
et al. \cite{wo94}). If the r-process originates from supernovae, a specific
progenitor mass dependence has to be assumed in order to reproduce the r-process
abundances in low metallicity stars as a function of [Fe/H] (Mathews et
al. \cite{ma92}; Wheeler et al. \cite{wh98}). Such a ``hypothetical'' r-process
yield curve has been constructed by Tsujimoto \& Shigeyama (\cite{ts98}), in
agreement with ideas of Ishimaru \& Wanajo (\cite{is99}) and Travaglio et
al. (\cite{tr99}), and is used in the present galactic evolution calculation.
However, we should keep in mind that present-day supernova models have
difficulties to reproduce the entropies required for such abundance calculations.
In addition, they could exhibit the incorrect abundance features of lighter
r-process nuclei (Freiburghaus et al. \cite{fr99a}), we know by now that at least
two r-process sources have to contribute to the solar r-process abundances
(Wasserburg et al. \cite{wa96}; Cowan et al. \cite{co99}), and that possible other
sources exist (Freiburghaus et al. \cite{fr99b}). A larger scatter in the r/Fe
ratio in low metallicity stars than predicted by the constructed supernova yields
would also indicate the need of such another r-process source.


\section{Observational Data}
\label{obsdat}

\begin{table*}
  \caption[]{Reference list of the observational data. Elements marked with
             ``c'' were adjusted to the same solar abundance scale by S.~G. Ryan
             (Ryan et al. \cite{ry96} and private communication), unaltered
             observations are marked with ``x''. Unmarked entries were not
             observed in the corresponding study. Newer publications (after 1995)
             are listed separately.}
  \begin{tabular*}{\hsize}[]{l@{\extracolsep\fill}ccccccccc}
    \hline
    \noalign{\smallskip}
      Author                                & O & Mg& Si& Ca& Cr& Mn& Fe& Ni& Eu\\
    \noalign{\smallskip}
    \hline
    \noalign{\smallskip}
      Gratton (\cite{gr89})                 &   &   &   &   &   & c & c &   &   \\
      Magain (\cite{ma89})                  &   & c &   & c & c &   & c &   & x \\
      Molaro \& Bonifacio (\cite{mb90})     &   & c &   & c & c &   & c &   &   \\
      Molaro \& Castelli (\cite{mc90})      &   & c & c & c & c & c & c & c &   \\
      Peterson et al. (\cite{pe90})         & x & c & c & c & c & c & c & c &   \\
      Zhao \& Magain (\cite{zh90})          &   &   &   & c & c &   & c & c &   \\
      Bessell et al. (\cite{be91})          & x &   &   &   &   &   & x &   &   \\
      Gratton \& Sneden (\cite{gr91a})      &   &   & c & c & c &   & c & c &   \\
      Gratton \& Sneden (\cite{gr91b})      &   &   &   &   &   &   & x &   & x \\
      Ryan et al. (\cite{ry91})             & x & c & c & c & c & c & c & c & x \\
      Spiesman \& Wallerstein (\cite{sw91}) & x &   &   &   &   &   & x &   &   \\
      Spite \& Spite (\cite{sp91})          & x &   &   &   &   &   & x &   &   \\
      Francois et al. (\cite{fr93})         &   &   &   &   &   &   & x &   & x \\
      Norris et al. (\cite{no93})           &   & c &   & c & c & c & c & c &   \\
      Beveridge \& Sneden (\cite{be94})     & x & x & x & x & x & x & x & x & x \\
      King (\cite{ki94})                    & x &   &   &   &   &   & x &   &   \\
      Nissen et al. (\cite{ni94})           & x & x &   & x & x &   & x &   &   \\
      Primas et al. (\cite{pr94})           &   & c & c & c & c & c & c &   &   \\
      Sneden et al. (\cite{sn94})           &   & x & x & x & x & x & x &   & x \\
    \noalign{\smallskip}
    \hline
    \noalign{\smallskip}
      Fuhrmann et al. (\cite{fu95})         & x &   &   &   &   &   & x &   &   \\
      McWilliam et al. (\cite{mw95})        &   & c & c & c & c & c & c & c & x \\
      Balachandran \& Carney (\cite{ba96})  & x &   &   &   &   &   & x &   &   \\
      Ryan et al. (\cite{ry96})             &   & c & c & c & c & c & c & c & c \\
      Israelian et al. (\cite{is98})        & x &   &   &   &   &   & x &   &   \\
      Jehin et al. (\cite{je99})            &   & x &   & x & x &   & x & x & x \\
    \noalign{\smallskip}
    \hline
  \end{tabular*}
  \label{obs}
\end{table*}

As shown in Table~\ref{obs}, the observational data were selected from various
high-resolution studies. Elements marked with ``c'' were adjusted to the same
solar abundance scale by S.~G. Ryan (Ryan et al. \cite{ry96} and private
communication), entries marked with ``x'' remained unaltered. Unmarked entries
were not observed in the corresponding study. All observations of very metal-poor
stars published after 1989 were taken into account. In the case where multiple
observations of stars exist, we used the most recent data. If they were published
in the same year or \emph{after} 1995, the values were averaged.

Typical abundance errors given for these observations are about 0.1 dex (see Ryan
et al. \cite{ry96}). These may not fully account for systematic errors, which
could, e.g., be caused by the choice of the employed stellar atmosphere models and
parameters, or by the employed solar abundance values. If there are systematic
offsets between different subsamples, this could enhance the scatter in the
combined sample.

Special attention has to be paid to Cr, Mn and O. In the case of Cr and Mn, data
published after 1995 show a decrease in the [Cr/Fe] and [Mn/Fe] ratios for lower
metallicities. Such trends are not present in older observations. Regardless of
this, it is not possible to reproduce metallicity-dependent trends with our model,
since we only use the stellar yields of Thielemann et al. (\cite{th96}), which
assume constant solar progenitor star metallicity. A possible explanation of the
[Cr/Fe] and [Mn/Fe] trends is the dependency of stellar yields on the metallicity
of the progenitor star of a SN Type~II (Samland \cite{sa97}) or a progenitor mass
dependent mixing of SN~II yields with the ISM (Nakamura et al. \cite{na99}).

In the data set of Israelian et al. (\cite{is98}), the [O/Fe] ratio increases at
lower metallicities. This behaviour is apparently different from previous
determinations and is not yet understood, but almost identical results were
published by a different group of observers (Boesgaard et al. \cite{bo99}). The
$\alpha$-elements oxygen and magnesium are produced mainly during the hydrostatic
burning phase of a high mass star and are only slightly affected by the actual
explosion. Furthermore, their yields depend in almost the same way on the mass of
the progenitor star (Thielemann et al. \cite{th96}; Woosley \& Weaver
\cite{ww95}). Thus observations of Mg for the objects discussed in the O abundance
determinations by Israelian et al. (\cite{is98}) and Boesgaard et al.
(\cite{bo99}) would be very interesting.


\section{Results}
\label{results}

\subsection{Chemical Mixing of the Halo ISM}
\label{mixing}

Starting with an ISM of primordial abundances, a first generation of ultra
metal-poor stars is formed in the model. After the first high-mass stars exploded
as SNe of Type~II, the halo ISM is dominated by local inhomogeneities, since the
SN events are spatially well separated and no mixing has yet occurred. In a very
metal-poor medium, a single SN event heavily influences its surroundings, so that
its remnant shows the element abundance pattern produced by that particular
core-collapse SN. Stars born out of this enriched material therefore inherit the
same abundances. Since SNe of different progenitor masses have different stellar
yields, stars formed out of an incompletely mixed ISM show a great diversity in
their element abundances.

\begin{figure*}
 \resizebox{\hsize}{!}{\includegraphics{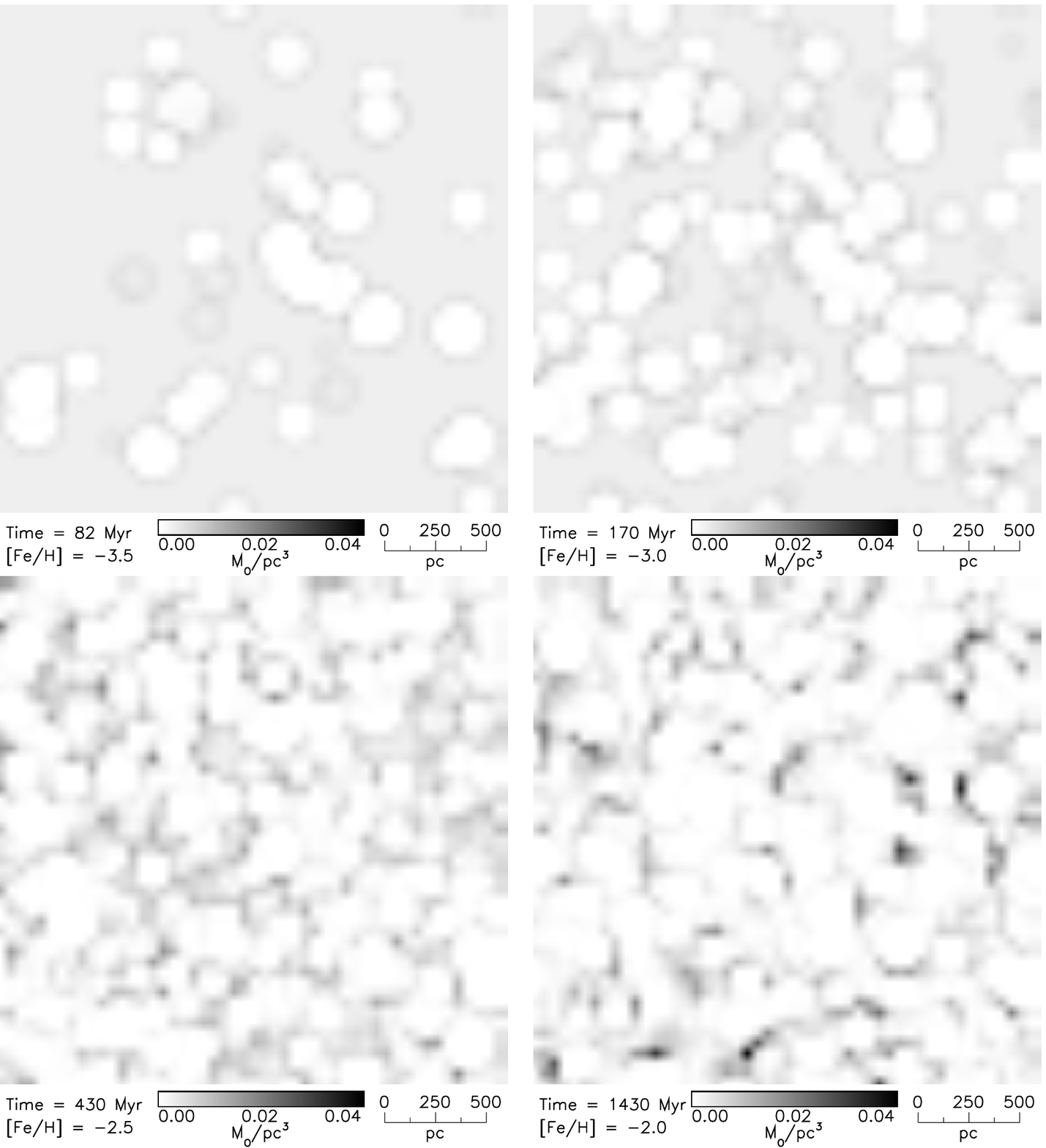}}
 \vspace{2em}
 \caption{Cut through the computed volume, showing the density distribution of the
          halo ISM during the transition from the unmixed to the well mixed
          stage (see text for details).}
 \label{density}
\end{figure*}

Fig.~\ref{density} shows a cut through the computed volume, giving the density
distribution of the halo ISM at four different times, ranging from the unmixed to
the well mixed stage. Each panel shows a different mean metallicity and has a
lateral length of 2.5 kpc. The density of the ISM ranges from $6 \cdot 10^{-7} \,
\mathrm{M}_{\sun}/pc^3$ in the inner part of a remnant to $0.04 \,
\mathrm{M}_{\sun} / pc^3$ in the densest clouds, which is about the gas density of
the solar neighbourhood (Binney \& Tremaine \cite{bt87}).

\emph{Upper left:} After 82 Myr (see below for the scaling of the time units with
the assumed SFR), the mean metallicity of the halo ISM is [Fe/H] $= -3.5$. Most of
the volume was not yet affected by SN events, which can be seen as bright patches
in the otherwise homogeneously distributed ISM. In the inner part of the remnants
most of the gas has been swept up and condensed in thin shells, which show up as
dark, ring-like structures. The regions influenced are well separated and no or
only slight mixing on a local scale has taken place. The halo ISM has to be
considered unmixed and is dominated by local inhomogeneities. Stars forming in the
neighbourhood of such inhomogeneities show an abundance pattern which is
determined by the ejecta of a single SN or a mixture of at most two to three SN
events.

\emph{Upper right:} Beginning of the transition from the unmixed to the well mixed
halo ISM, after 170 Myr at a mean metallicity of [Fe/H] $= -3.0$. The separation
of single SN remnants has become smaller and a higher number of remnants may
overlap. The abundance pattern in overlapping shells still shows a great diversity
but is closer to the IMF averaged element abundance than at the completely unmixed
stage.

\emph{Lower left:} After 430 Myr, at a mean metallicity of [Fe/H] $= -2.5$, enough
SN events have occurred to pollute the mass of the whole ISM twice. Nevertheless
there are still cells which were not influenced by a S, showing therefore
primordial abundances.

\emph{Lower right:} After 1430 Myr, a mean metallicity of [Fe/H] $= -2.0$ is
reached. At this time, no patches with primordial abundances exist and the ISM
starts to be chemically well mixed. The once homogeneous, primordial medium has
now a lumpy structure with large density fluctuations and shows mainly an IMF
averaged abundance pattern.

\medskip

\begin{figure*}
 \resizebox{0.85\hsize}{!}{\includegraphics{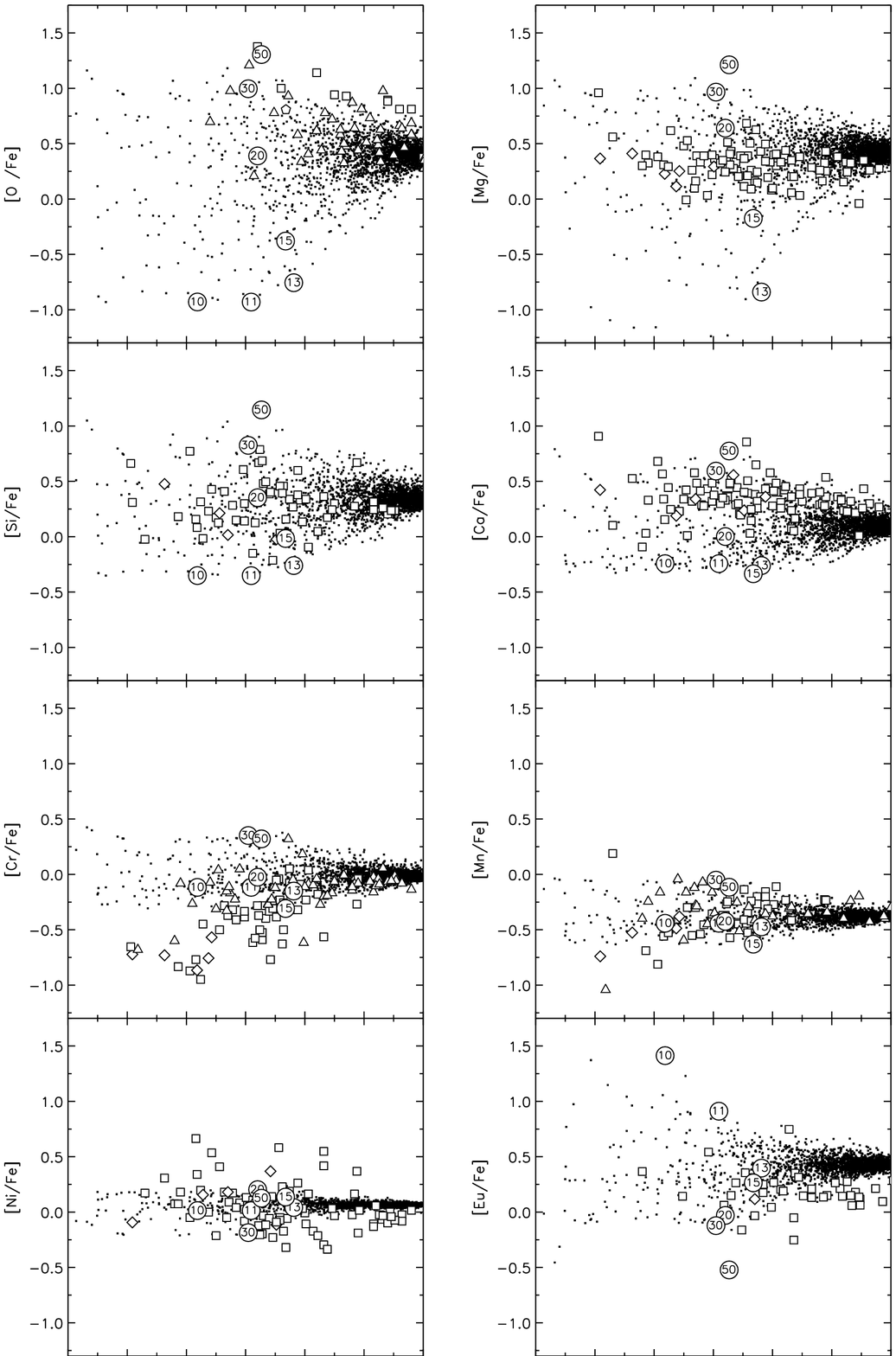}}
 \vspace{1.5em}
 \caption{Element-to-iron ratio [El/Fe] of O, Mg, Si, Ca, Cr, Mn, Ni and Eu. Open
          circles depict [El/Fe] ratios of SN~II models of the given progenitor
          mass. Small filled squares represent model stars, open symbols show
          observed stars.}
 \label{scatter}
\end{figure*}

Our calculations confirm that the incomplete mixing of the ISM during the halo
formation plays a significant r\^ole in the early enrichment of the metal-poor
gas. This can be seen in the [El/Fe] ratios of the considered elements, shown in
Fig.~\ref{scatter}. The small, filled squares show the [El/Fe] ratio of single
model-stars. For comparison, observed metal-poor stars are represented by open
squares. In the case of O, Cr and Mn, observations taken before 1995 were marked
with open triangles. This was done to highlight possible trends in the [El/Fe]
ratio of these elements. If multiple observations of a metal-poor star occurred
and the abundances had to be averaged, open diamonds were used (see
Sect. \ref{obsdat} for details).

The numbered circles show the [El/Fe] and [Fe/H] ratios of a single SN~II with the
indicated mass. The [Fe/H] ratio in the shell of the SN remnants are determined by
the mass of the exploding star and the $5 \times 10^4 \, \mathrm{M}_{\sun}$ of gas
that are swept up by the shock front. This ratio represents the maximal abundance
a \emph{single} SN~II can produce and determines the position of the open circles
on the [Fe/H] axis in Fig.~\ref{scatter}. If the swept up material subsequently
mixes with the surrounding medium, this abundance decreases and stars with lower
[Fe/H] abundance can be formed. On the other hand, the [El/Fe] ratio is determined
solely by the stellar yields of a SN~II and is independent of the mixing mass. For
the uncertainties of Fe-ejecta and [El/Fe] ratios in individual SNe~II see the
discussion in Sect. \ref{nucleo}.

We divide the chemical enrichment process in the early evolution of the halo ISM
into different enrichment stages: at metallicities [Fe/H] $<-3.0$, the ISM is
completely unmixed and dominated by local inhomogeneities, originating from SN~II
events. At about [Fe/H] $> -2.0$ the halo ISM shows an IMF averaged abundance
pattern and has to be considered chemically well mixed. The continuous transition
between these two phases is marked by incomplete mixing which gradually becomes
better, as more and more high-mass stars explode as SN~II.

The different phases in the enrichment of the halo ISM seen in Fig.~\ref{scatter}
can be distinguished in all [El/Fe] -- [Fe/H] plots. At very low metallicities
([Fe/H] $< -3.0$), only a few stars exist, showing a considerable spread in their
[El/Fe] ratios, ranging from 0.5 dex in the case of Ni to more than 2 dex in the
cases of O, Mg and Eu. At this stage, the scatter of the model stars is given by
the spread in metallicities of the SN models. At the end of this early phase, less
than 0.5\% of the total halo ISM mass has been transformed into stars.

At a metallicity of $-3.0 <$ [Fe/H] $< -2.0$, the SN remnants start to overlap and
a first, incomplete mixing occurs. New stars form out of material which was
influenced by several SNe of different masses. Therefore, they do not show the
typical abundance pattern of a single SN, but show an average of the SNe which
contributed to the enrichment of the local ISM. The spread in the metallicities
gradually decreases from [Fe/H] $= -3.0$ to $-2.0$, reflecting the ongoing mixing
process as more and more SNe pollute the ISM. At the beginning of the well-mixed
phase, star-formation has consumed about 2.5\% of the total halo ISM mass.

This late phase is characterized by a well mixed ISM and begins at [Fe/H] $>
-2.0$, where the abundance scatter in the model is reduced to a third of its
initial value. At this stage, the whole volume considered was influenced several
times by SN events. This leads to an IMF averaged [El/Fe] ratio in the ISM which
is the same as predicted by simple 1-zone models and which is observed in stars
with metallicity [Fe/H] $> -2.0$. Even in an enriched medium, however, a SN event
will still have an influence on its neighbourhood, although the change in the
abundance pattern will not be as prominent as in a very metal-poor medium. This
explains the fact, that even in the well-mixed case, the [El/Fe] ratios show a
certain dispersion. At [Fe/H] $=-1.0$, about 8\% of the total halo ISM mass has
been used to form stars.

The onset of SNe of Type~Ia marks the beginning of a third phase in the chemical
enrichment history of the galaxy, but we will not consider this phase any
further. Note, that the mean [El/Fe] ratios at [Fe/H] $= -1.0$ in our model do not
have to be equal to zero (by definition the solar metallicity), since we have
neglected the influence of SN~Ia events.

\medskip

To quantify the enrichment of the ISM we introduce the concept of the polluted
mass $M_{\mathrm{poll}}$ in a unit volume at time $\tau$. It is defined as the
total mass which gets polluted by $N_{\mathrm{SN}}$ \emph{isolated} SNe, where
$N_{\mathrm{SN}}$ is the number of SNe in the unit volume that occurred during the
elapsed time $\tau$. For a constant mixing-mass $M_{\mathrm{sw}}$ swept up by a SN
event, $M_{{\rm poll}}=N_{\mathrm{SN}} \cdot M_{\mathrm{sw}}$. Since the polluted
mass is directly proportional to the number of SNe, it can become larger than the
total ISM mass in the unit volume, $M_{\mathrm{tot}}$. Furthermore, the pollution
factor $f_{{\rm poll}}$ is defined as the ratio $M_{\mathrm{poll}} /
M_{\mathrm{tot}}$ and only depends on $N_{\mathrm{SN}}$, for fixed
$M_{\mathrm{sw}}$ and $M_{\mathrm{tot}}$. When
$f_{\mathrm{poll}}=1$, enough SNe have contributed to the chemical enrichment to
theoretically pollute the entire ISM in the unit volume, even though there still
may be patches of material which were not yet affected by any SN. A higher
pollution factor results in a better mixing of the halo ISM, decreasing the local
abundance differences and the amount of material with primordial abundances.
Therefore, the ratio $M_{\mathrm{sw}}/M_{\mathrm{tot}}$ determines the mixing
efficiency in our model, i.e. how many SNe in the unit volume are needed to reach
a certain value of $f_{\mathrm{poll}}$. Given $f_{\mathrm{poll}}$,
$M_{\mathrm{sw}}$ and the mean, IMF integrated iron yield $\left< M_{\mathrm{Fe}}
\right>$ of a typical SN~II, the mean metallicity of the ISM then is determined by
\begin{eqnarray*}
\mathrm{[Fe/H]} = \log \frac{N_{\mathrm{SN}} \cdot
\left< M_{\mathrm{Fe}} \right>}{M_{\mathrm{tot}}} - C
= \log \frac{f_{\mathrm{poll}} \cdot
\left< M_{\mathrm{Fe}} \right>}{M_{\mathrm{sw}}} - C,
\end{eqnarray*}
where $C$ is the solar iron abundance.

The local inhomogeneities of the ISM begin to disappear when most of the gaseous
SN remnants start to overlap. This is the case when more or less every cloud in
the halo was influenced at least once by a SN event, i.e. the pollution factor is
about equal to one. With the adopted mixing mass of $M_{\mathrm{sw}} = 5 \times
10^4 \, \mathrm{M}_{\sun}$ this is the case at [Fe/H] $\approx -2.8$. This
metallicity gives an upper limit for the end of the early phase and the beginning
of the transition to the second, well-mixed enrichment phase.

\begin{table}
  \caption[]{Pollution factor and SN Type~II frequency.}
  \begin{tabular*}{\hsize}[]{l@{\extracolsep\fill}ccc}
    \hline
    \noalign{\smallskip}
      [Fe/H]  & $f_{\mathrm{poll}}$ & $N_{\mathrm{SN}}$ [kpc$^{-3}$] & 
      $\tau$ [Myr] \\
    \noalign{\smallskip}
    \hline
    \noalign{\smallskip}
      -3.5 & 0.2 &  27 &   82 \\
      -3.0 & 0.7 &  83 &  170 \\
      -2.5 & 2.1 & 260 &  430 \\
      -2.0 & 6.5 & 810 & 1430 \\
    \noalign{\smallskip}
    \hline
  \end{tabular*}
  \label{snnr}
\end{table}

Table~\ref{snnr} shows the pollution factor needed to reach the mean metallicities
shown in the panels of Fig.~\ref{density}. Also shown are the corresponding SN
frequency $N_{\mathrm{SN}}$ and elapsed time $\tau$, which depend on our model
parameters. Here, the SN frequency is defined as the number of SNe per kpc$^3$ and
depends on the total ISM mass in the unit volume, whereas the elapsed time scales
with the average SFR as
\begin{eqnarray*}
\tau^{\prime} = \tau \cdot \left( \frac{ \left< \mathrm{SFR} \right>}{1.06 \cdot
10^{-4}} \right)^{-1},
\end{eqnarray*}
where $1.06 \cdot 10^{-4} \, \mathrm{M}_{\sun} \, \mathrm{yr}^{-1} \,
\mathrm{kpc}^{-3}$ is the mean SFR in the unit volume in our model. Note that the
evolution of the abundance ratios as a function of [Fe/H] is independent of the
star formation timescale and the SFR specified in the model.

\subsection{Comparison with Observations}

\begin{figure}
 \resizebox{\hsize}{!}{\includegraphics{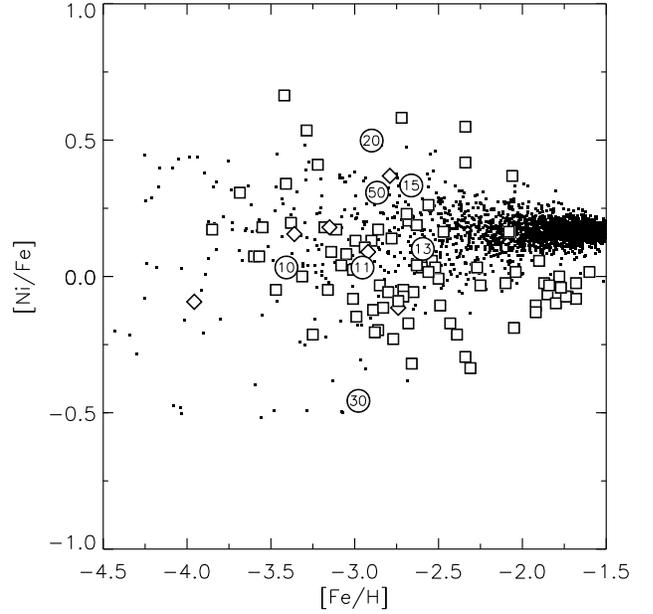}}
 \caption{Normalized nickel-to-iron ratio [Ni/Fe] as function of metallicity
          [Fe/H]. Symbols are the same as in Fig.~\ref{scatter}. The scatter
          of the model-stars and the halo stars was normalized to unity to 
          highlight the enrichment phases of the halo ISM.}
 \label{ninormal}
\end{figure}

The scatter in the [El/Fe] ratios of the model stars as a function of [Fe/H] shows
the same general trend for every element considered, independent of the individual
stellar yields. The inhomogeneous mixing of the very metal-poor halo ISM at [Fe/H]
$< -3.0$ leads to a scatter in the [El/Fe] ratios of up to 1 dex. This scatter
continuously decreases for higher metallicities, reflecting the ongoing mixing of
the ISM. At [Fe/H] $> -2.0$ the model stars show an IMF averaged abundance pattern
with an intrinsic scatter of about 0.1 to 0.2 dex. This behaviour matches the
general trend of the observations well, as can be seen in Fig.~\ref{scatter}. The
observations also show a large scatter at low metallicities which again decreases
for higher [Fe/H], with some exceptions, however: the iron-group elements Cr and
Mn show a strong decrease in the [Cr/Fe] and [Mn/Fe] ratio for lower
metallicities. This behaviour can not be reproduced with our adopted
metallicity-independent stellar yields and the progenitor-independent mixing mass.

Compared to the observations, the distribution of [Ni/Fe] ratios of the model
stars in Fig.~\ref{scatter} shows a scatter that is much too small. This is most
likely due to the choice of mass cuts in the SNe~II models, which have been set
with the aim to reproduce the average solar [Ni/Fe] ratio. Thielemann et
al. (\cite{th96}) discuss in detail that large variations can easily occur. See
also the discussion in Sect.~\ref{nucleo}. We therefore now want to investigate
whether the sequence of enrichment stages seen in our model is similar to the
observed evolution of abundance ratios even in cases when the employed yields may
be incorrect.

\begin{figure}
 \resizebox{\hsize}{!}{\includegraphics{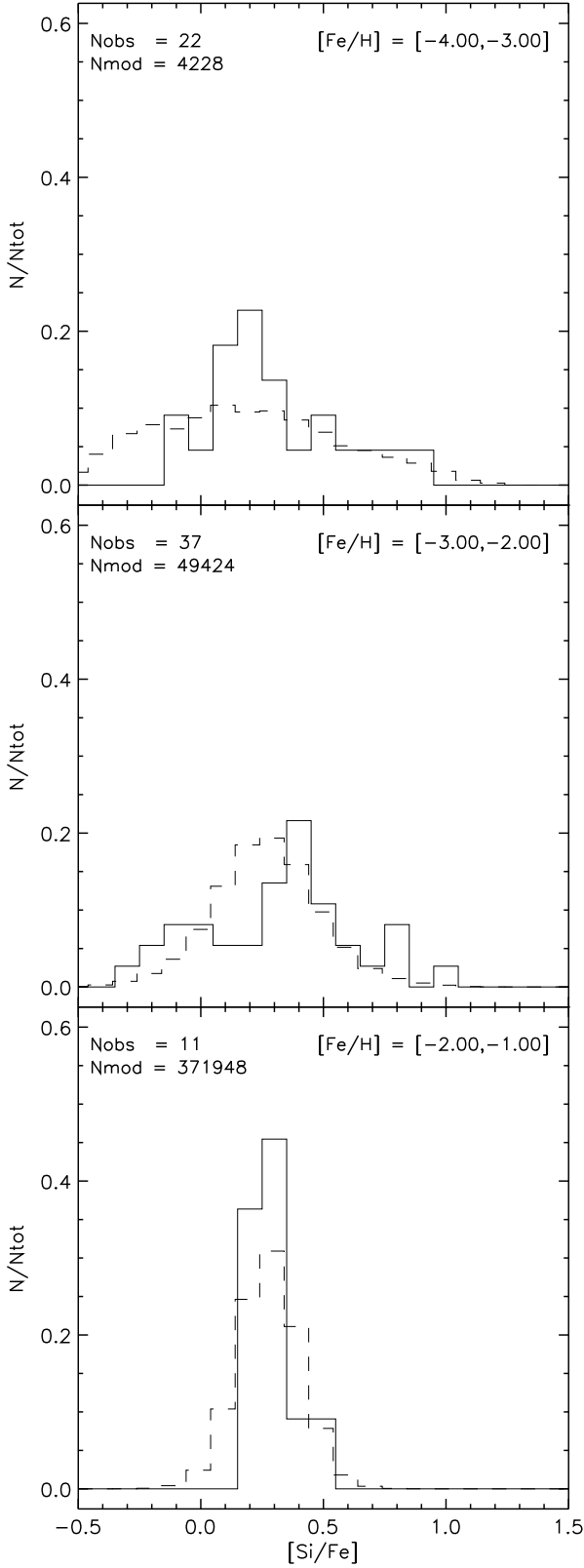}}
 \vspace{1.5em}
 \caption{Relative frequency of stars normalized to unity in [Si/Fe] bins for
          three different metallicity ranges (see text). The solid line shows
          observational data, the dashed line the model stars. The number of
          included stars is given in the upper left corner.}
 \label{sihist}
\end{figure}

To this end we have normalized the scatter in [Ni/Fe] of the model-stars at low
[Fe/H] to unity, and have similarly renormalized the range of values for the
observed stellar [Ni/Fe] ratios to one. The mean values of both distributions were
left unchanged. The resulting renormalized distributions are shown in
Fig.~\ref{ninormal}. The remarkably good agreement of both distributions after
this procedure indicates that the enrichment history of the halo ISM implied by
the model is consistent with the data, even though the employed Ni yields are
not. Based on similar comparisons, we conclude that the abundance ratio data of
most elements except Mn and Cr are consistent with the predicted enrichment
history, and the scatter plots in Fig.~\ref{scatter} can thus be used to compare
the range of the theoretically predicted nucleosynthesis yields with observations.

\vspace{1em}

To describe the transition from the metal-poor, unmixed to the enriched, well
mixed ISM more quantitatively, the relative frequency of stars at a given [El/Fe]
ratio has been analysed for the different enrichment phases. In the case of
silicon, this detailed enrichment history is shown in Fig.~\ref{sihist}. The
different enrichment phases from [Fe/H] $< -3.0$, $-3.0 <$ [Fe/H] $< -2.0$ and
[Fe/H] $> -2.0$ are given in the panels from top to bottom. The solid line shows
the relative frequency of observed halo stars per [Si/Fe] bin for each enrichment
phase and the dashed line the relative frequency of computed model stars per
bin. To account for the effect of observational errors on our data, we added a
random, normally distributed error with standard deviation 0.1 dex in the [El/Fe]
and [Fe/H] ratios to the model stars. The bin size in the [Si/Fe] ratio is 0.1 dex
for observed and computed stars, while the position of the histogram for the model
stars is shifted by 0.01 dex to the left for better visibility. The total number
of stars included in the plot is given in the upper left corner of each panel,
where $N_{\mathrm{obs}}$ and $N_{\mathrm{mod}}$ are the number of observed stars
and of model stars, respectively.

In the upper panel, the distributions of both the 22 observed and the 4226 model
stars show a spread in the [Si/Fe] ratio of more than one dex. The distribution of
the model stars shows two wide, protruding wings and a faint peak at [Si/Fe]
$\approx 0.2$. The ``right'' wing shows a shallow rise from [Si/Fe] $\approx 1.1$
to the peak. The ``left'' wing is not as extended and shows a rather steep cutoff
at [Si/Fe] $\approx -0.3$. This asymmetry is due to the nucleosynthesis models of
core-collapse SNe, which show a more or less constant value of [Si/Fe] $\approx
-0.3$ for progenitor masses in the range of $10-13 \, \mathrm{M}_{\sun}$, as can
be seen in Fig.~\ref{scatter}. The distribution of the halo stars peaks at the
same location as the model stars but extends only down to [Si/Fe] $\approx
-0.1$. We attribute this to the poor statistic of the data set, since this gap is
filled in the middle panel.

The middle panel of Fig.~\ref{sihist} shows the same distribution for the
intermediate mixing stage of the ISM. The distribution of the model stars now has
smaller wings, and peaks at [Si/Fe] $\approx 0.3$. It is still broader than 1 dex,
but the majority of the stars fall near the IMF averaged [Si/Fe] ratio. The
prominent peak is caused by the already well-mixed regions, whereas the broad
distribution shows that the halo ISM is still chemically inhomogeneous. The peak
of the observational sample has shifted by about 0.2 dex to the right and lies now
at [Si/Fe] $\approx 0.4$. Compared to the prediction of the model, the relative
frequency of the halo stars is too high in the wings of the distribution and too
low to the left of the peak.

The lower panel shows the late stage, where the halo ISM is well mixed. The broad
wings have completely disappeared and only the very prominent peak at the IMF
averaged value remains. The distributions of the 11 observed stars and the
370\,000 model stars are in good agreement. At this metallicity no SN of Type~Ia
should have polluted the by now well mixed ISM and the metal abundance is high
enough to restrict the impact of single SN~II events on the ISM.

\begin{figure}
 \resizebox{\hsize}{!}{\includegraphics{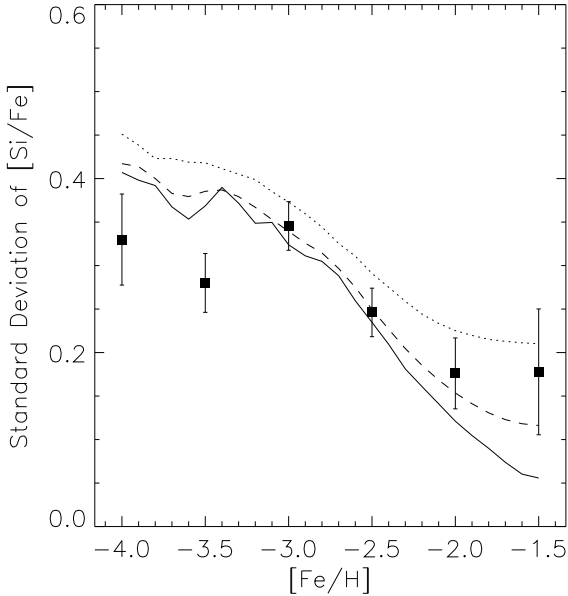}}
 \caption{Scatter in [Si/Fe] of the model and observed stars. The solid line
          gives the scatter of the model stars, the dashed and dotted lines show
          the scatter of the model stars, folded with an error of 0.1 and 0.2 dex.
          Filled squares give the standard deviation of the observed stars
          (see text).}
 \label{sidis}
\end{figure}

\vspace{1em}

The most prominent feature which characterizes the different enrichment phases, is
the intrinsic scatter in the abundances of metal-poor stars. This can be seen in
Fig.~\ref{sidis}, which shows the standard deviation of [Si/Fe] as a function of
metallicity [Fe/H] for the model and the halo stars. The bin size used to compute
the standard deviation was 0.1 dex in metallicity. The solid line shows the
scatter of the unmodified model stars. The influence of observational errors on
our data was simulated by adding a random, normally distributed error with
standard deviation 0.1 and 0.2 dex in both [Si/Fe] and [Fe/H]. The resulting
scatter in dependence of metallicity is given by the dashed and dotted lines. In
the range of $-4.0 <$ [Fe/H] $< -3.0$ the scatter has a more or less constant
value of approximately 0.4 dex. It declines rather steeply in the range $-3.0 <$
[Fe/H] $< -2.0$ and levels off again at metallicities higher than $-2.0$,
depending on the assumed observational errors of 0.0, 0.1 or 0.2 dex. These curves
show that for errors in this range the scatter at low metallicities is dominated
by the intrinsic differences in the element abundances of single stars.

For comparison, the scatter in the [Si/Fe] ratio of observed halo stars is
represented by filled squares. The observations were binned with a bin size of 0.5
dex to compute the standard deviations. To estimate the reliability of their
scatter in [Si/Fe], we built several new data sets by adding a normally
distributed random error with standard deviation 0.1 dex to the [Si/Fe] ratio and
the metallicity of the stars. For each new data set, the standard deviation in the
different bins was computed. The standard deviation for the results from these
artificial data sets is given in the plot as 1-$\sigma$ error-bars. The scatter of
the observed abundance ratios shows nicely the features already seen in the curves
for the model stars. At the first stage of the enrichment, it is approximately
constant, followed by a steady decline in the intermediate mixing phase. At higher
metallicities, the scatter levels off again.

Since the scatter in [Si/Fe] at [Fe/H] $=-4.0$ is about $0.4$ dex the
observational errors have little influence on the analysis at these low
metallicities, unless unknown systematic or confusion errors were large enough to
inflate the scatter at [Fe/H] $=-4.0$ to also about $0.4$ dex. On the other hand,
observational errors do dominate the scatter at metallicities [Fe/H] $> -2.0$,
when the halo ISM is well mixed and the intrinsic scatter of the stars is
negligible compared to the observational errors.

\subsection{Individual Elements and Nucleosynthesis}
\label{individual}

\emph{Oxygen \& Magnesium:} As expected, the IMF averaged [El/Fe] ratios for O and
Mg reproduce the mean abundance of the observed metal-poor stars nicely. The
[O/Fe] ratio seems to be slightly too low, whereas [Mg/Fe] is slightly too high,
but both deviations are smaller than 0.1 dex. No trend in the observational data
of Mg can be seen and a trend in O only becomes visible if the observations of
Israelian et al. (\cite{is98}) are considered.

An important fact is that the scatter in the data, although increasing at lower
metallicities, does not match the large scatter of more than two dex predicted by
the stellar yields. Since no other mixing effects than the overlapping of SN
remnants are included in our model, the expected scatter is determined by the
nucleosynthesis yields. If gas flows and the random motion of stars in the halo
accelerated the chemical mixing, a smaller scatter in the model data would be
expected. Even then, the fact that the observed stars only show [O/Fe] or [Mg/Fe]
ratios corresponding to the stellar yields produced by $18-50 \,
\mathrm{M}_{\sun}$ SNe would remain unexplained.

If we assume a top-heavy IMF which favoured high-mass SNe, this problem could be
solved. However, the abundance pattern of the other elements should then also
reflect this, which is not the case. This leaves us with two explanations: Either
the stellar yields of O \emph{and} Mg or of Fe are incorrect (or both).

Since the exact location of the mass cut is not known, the actual Fe yields are
not very well determined (Woosley \& Weaver \cite{ww95}; Thielemann et al.
\cite{th96}) and direct observational information which links a progenitor to an
ejected Fe mass is very limited, maybe with the exception of SN 1987A and 1993J.
Otherwise only the IMF integrated Fe-yields are constrained, but not necessarily
their progenitor mass dependence (declining, rising, or with a maximum, see
Nakamura et al. 1999 and Sect.~\ref{nucleo}). In order to attain a fit to the
observational data within our evolution model we would need to decrease the Fe
yields of the $13$ or $15 \, \mathrm{M}_{\sun}$ stars by a factor of six. Without
adjusting Fe-yields of the more massive stars, and assuming a standard Salpeter
IMF, this would increase the IMF averaged [O/Fe] and [Mg/Fe] ratio by about 0.3
dex (a factor of two) and would therefore result in a much too high IMF averaged
value. Equally, every other abundance ratio would be affected by this change.

On the other hand, the stellar yields of the $\alpha$-elements O and Mg could be
too low for the 13 and $15 \, \mathrm{M}_{\sun}$ progenitors. The abundances of O
and Mg are mainly determined in the hydrostatic burning phases and do not depend
heavily on the explosion mechanism. Changing the stellar yields of O and Mg for
the 13 and $15 \, \mathrm{M}_{\sun}$ progenitor stars would therefore require to
adjust the existing stellar models, which suffer from uncertainties in the theory
of convection and the treatment of rotation.

\emph{Silicon:} The IMF averaged [Si/Fe] ratio and the scatter predicted by the
stellar yields fit the observations perfectly. The decrease in the scatter for
higher metallicities and therefore the different enrichment phases of the halo ISM
are clearly visible. At [Fe/H] $< -3.0$, the scatter in the model points reflects
the scatter predicted by the stellar yields. During the transition from the
unmixed to the well mixed ISM, in the range $-3.0 <$ [Fe/H] $< -2.0$, the scatter
decreases steadily and leads to the IMF averaged [Si/Fe] ratio, which is reached
at [Fe/H] $> -2.0$.

\emph{Calcium:} The IMF averaged [Ca/Fe] ratio is about 0.2 dex lower than the
observed mean for metal-poor stars. If the model data is shifted by this value to
reproduce the mean of the observational data, the scatter for very metal-poor
stars and the transition to the less metal-poor stars fits the data well. Note
that, contrary to the other $\alpha$-elements, the stellar yields of Ca are no
longer approximately proportional to the mass of the progenitor. This behaviour
becomes more pronounced in the cases of Cr, Mn and Ni.

\emph{Chromium \& Manganese:} The iron-peak elements Cr and Mn are both produced
mainly during explosive silicon burning and show an almost identical, complicated
dependence on the mass of the progenitor. The IMF averaged [El/Fe] ratio
reproduces the observations of the less metal-poor stars well. The scatter
predicted by the stellar yields is only about 0.8 dex and is not as large as for
the $\alpha$-elements.

A notable feature of the observations is the decrease of the [Cr/Fe] and [Mn/Fe]
ratios for lower metallicities, seen in the newer data. If these trends are real,
they can not be reproduced by metallicity-independent yields unless one assumes a
progenitor mass dependent amount of mixing with the interstellar medium (Nakamura
et al. \cite{na99}).

The upper limits of the [El/Fe] ratios of Cr and Mn are given by the stellar
yields of a $30 \, \mathrm{M}_{\sun}$ SN and correspond to the highest [Cr/Fe] and
[Mn/Fe] ratios seen in metal-poor stars, with the exception of the binary CS
22876-032 which shows an unusual high [Mn/Fe] ratio of 0.29 dex. Recent
high-signal-to-noise, high-resolution data result in a lower [Mn/Fe] value for
this star, placing it near the IMF averaged value (S.~G. Ryan, private
communication). On the other hand, the lower limits of the observed [Cr/Fe] and
[Mn/Fe] ratios do not correspond at all with those given by the stellar yields,
which show a ratio which is too high by up to 0.5 dex in the Thielemann et
al. (\cite{th96}) yields. Only a different choice of mass cuts as a function of
progenitor mass (Nakamura et al. \cite{na99}) or as a function of metallicity
would be able to rectify this.

\emph{Nickel:} The stellar yields of the most important iron-peak element besides
Fe completely fail to reproduce the observations. The scatter of the stellar
yields is only about 0.5 dex compared to about 1.2 dex seen in the observational
data. Compared to the mean of the observations, the IMF averaged abundance of the
model-stars is about 0.1 dex to high. The small scatter in the [Ni/Fe] ratio
originates from an almost constant ratio of Ni and Fe yields, which means that in
the models Ni is produced more or less proportional to iron. To reproduce the
scatter seen in the observations, Ni would have to depend on the progenitor mass
differently from Fe. This is a shortfall of the employed yields, adjusted to
reproduce the average solar [Ni/Fe] ratio with the choice of their mass
cuts. Varying neutron excess in the yields, however, can change the [Ni/Fe] ratio
drastically (Thielemann et al. \cite{th96} and Sect.~\ref{nucleo}).

\emph{Europium:} The r-process element Eu reproduces the scatter and the mean of
the observational data quite well. But compared to the $\alpha$-elements O, Mg, Si
and Ca its behaviour is very different. The highest [Eu/Fe] ratio is produced by
low mass SN and the lowest ratio by high mass SN, as required when one constructs
yields under the assumption that the r-process originated from SNe~II (see
Sect.~\ref{nucleo}). This is exactly the opposite to what is found for the
$\alpha$-elements. Therefore, a top-heavy IMF would lead to a steadily increasing
[Eu/Fe] ratio, since the most massive SNe will explode first (cf. \emph{Oxygen \&
Magnesium}).

\subsection{Age--Metallicity Relation}

\begin{figure}
 \resizebox{\hsize}{!}{\includegraphics{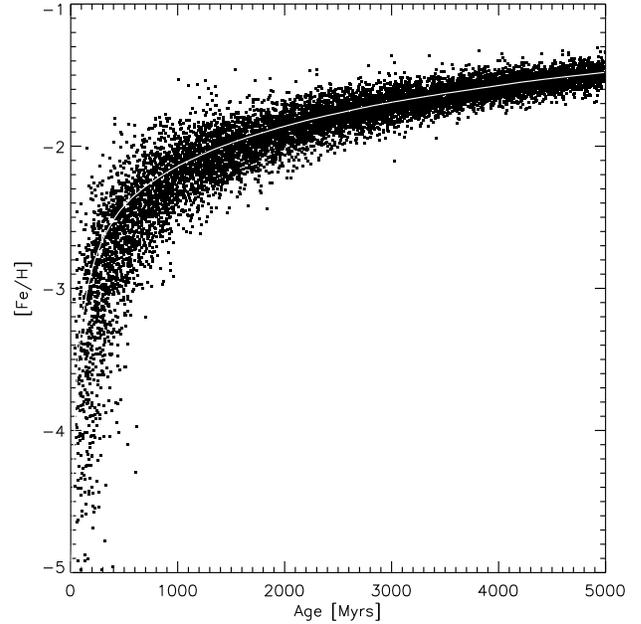}}
 \caption{Metallicity [Fe/H] vs. age of single model stars.}
 \label{agemetal}
\end{figure}

Common 1-zone chemical evolution models are based on the assumption that the
system is well mixed at all times. An important consequence of this assumption is
a monotonically increasing metallicity, which leads to a well defined
age--metallicity relation. Therefore, it is in principle possible to deduce the
age of a star if its metallicity is known. This basic assumption of 1-zone
chemical evolution models was dropped in our stochastic approach. Therefore, it is
not surprising that the well defined age--metallicity relation has to be replaced
by a statistical relation. In Fig.~\ref{agemetal}, the metallicity [Fe/H] of model
stars is plotted against the time of their formation. Model stars are represented
by small filled squares. For comparison the white line visible in the middle of
the black strip shows the mean age--metallicity relation, corresponding to the
relation given by a 1-zone model. As can be seen, there is no clear
age--metallicity relation at any time. Stars which were formed in the first 500
million years show a metallicity ranging from [Fe/H] $<-4.0$ up to [Fe/H] $>-2.0$
and in one extreme case up to [Fe/H] $\approx -1.5$. On the other hand, stars with
metallicity [Fe/H] $= -2.0$ could have formed at any time in the first $3 \times
10^9$ years. Taking these huge uncertainties into account, it is no longer
possible to speak of a well-defined age--metallicity relation.

While the mean [Fe/H] abundance increases about linearly with time, the scatter in
Fig.~\ref{agemetal} again reflects the different enrichment phases of the ISM. The
steep rise at early times marks the metal-poor and chemically inhomogeneous stage
of the halo ISM. Contrary to this first phase, where SN events dominated the
metal-poor ISM locally and the enrichment of isolated clouds could be very
efficient, the late stage is characterized by a well mixed ISM and therefore an
inefficient enrichment, which is reflected by the slow increase of the
age--metallicity ``relation'' at later times.

\subsection{Ultra Metal-poor Stars}
\label{ultra}

\begin{table}
  \caption[]{Top: Relative frequency of stars in the homogeneous intermediate
             resolution survey of Ryan \& Norris (\cite{rn91}), the combined
             high resolution data from Table~\ref{obs} and our model, binned with
             binsize 1 dex.

             Bottom: Absolute numbers. The last row gives the number of stars per
             bin which we expect to be present, if our model gives a fair
             representation of the halo metallicity distribution. The number of
             model-stars is normalized to the number of stars in the range $-4.0
             <$ [Fe/H] $< -3.0$ in the high resolution sample. No star was
             detected with confirmed [Fe/H] $< -4.0$, in contrast to the $5 \pm 2$
             stars predicted by the model.}
  \begin{tabular*}{\hsize}[]{l@{\extracolsep\fill}ccc}
    \hline
    \noalign{\smallskip}
      [Fe/H]  & $[-3.0,-2.0]$ & $[-4.0,-3.0]$ & $[-5.0,-4.0]$ \\
    \noalign{\smallskip}
    \hline
    \noalign{\smallskip}
      Ryan \& Norris  & 0.943 & 0.057 & 0.000 \\
      High resolution & 0.654 & 0.346 & 0.000 \\
      Model           & 0.865 & 0.115 & 0.020 \\
    \noalign{\smallskip}
    \hline
    \noalign{\smallskip}
      Ryan \& Norris  & $99$ & $ 6$ & $ 0$ \\
      High resolution & $53$ & $28$ & $ 0$ \\
      Expected        & $211 \pm 15$ & $28 \pm 5$ & $5 \pm 2$ \\
    \noalign{\smallskip}
    \hline
  \end{tabular*}
  \label{starnr}
\end{table}

\begin{table}
  \caption[]{Relative frequency of model stars, binned with binsize 0.2 dex.}
  \begin{tabular*}{\hsize}[]{l@{\extracolsep\fill}ccc}
    \hline
    \noalign{\smallskip}
      [Fe/H]  & $N/N_{tot}$ & $\log(N/N_{tot})$ \\
    \noalign{\smallskip}
    \hline
    \noalign{\smallskip}
     $[-5.0,-4.8]$ & $1.92 \cdot 10^{-03}$ & -2.72 \\
     $[-4.8,-4.6]$ & $2.37 \cdot 10^{-03}$ & -2.63 \\
     $[-4.6,-4.4]$ & $3.43 \cdot 10^{-03}$ & -2.46 \\
     $[-4.4,-4.2]$ & $5.10 \cdot 10^{-03}$ & -2.29 \\
     $[-4.2,-4.0]$ & $7.51 \cdot 10^{-03}$ & -2.12 \\
     $[-4.0,-3.8]$ & $1.00 \cdot 10^{-02}$ & -2.00 \\
     $[-3.8,-3.6]$ & $1.31 \cdot 10^{-02}$ & -1.88 \\
     $[-3.6,-3.4]$ & $1.93 \cdot 10^{-02}$ & -1.71 \\
     $[-3.4,-3.2]$ & $2.82 \cdot 10^{-02}$ & -1.55 \\
     $[-3.2,-3.0]$ & $4.41 \cdot 10^{-02}$ & -1.36 \\
     $[-3.0,-2.8]$ & $6.71 \cdot 10^{-02}$ & -1.17 \\
     $[-2.8,-2.6]$ & $1.02 \cdot 10^{-01}$ & -0.99 \\
     $[-2.6,-2.4]$ & $1.47 \cdot 10^{-01}$ & -0.83 \\
     $[-2.4,-2.2]$ & $2.11 \cdot 10^{-01}$ & -0.68 \\
     $[-2.2,-2.0]$ & $3.38 \cdot 10^{-01}$ & -0.47 \\
    \noalign{\smallskip}
    \hline
  \end{tabular*}
  \label{starnr2}
\end{table}

\begin{figure}
 \resizebox{\hsize}{!}{\includegraphics{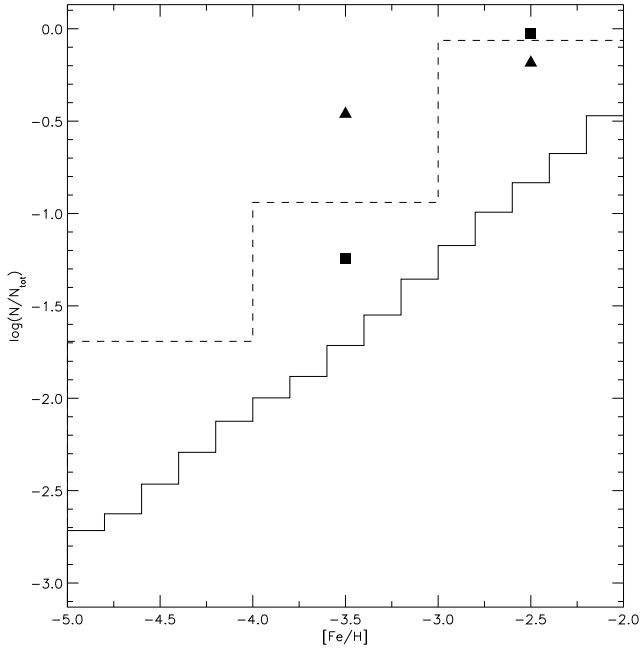}}
 \caption{Metallicity distribution in the model at the end of the calculation
          ([Fe/H]$=-1$). The number of stars with metallicity [Fe/H] is shown (a)
          in bins of 0.2 dex (b) in bins of 1.0 dex. In both cases the total
          number of stars is normalized to one. The solid squares show the
          intermediate resolution data of Ryan \& Norris (\cite{rn91}), the solid
          triangles the combined high resolution data from Table~\ref{obs}, both
          binned with bin size 1.0 dex and similarly normalized. There are no
          observed stars with [Fe/H] $< -4.0$. See Tables~\ref{starnr},
          \ref{starnr2} and text.}
 \label{number}
\end{figure}

From our calculations, we can deduce the number of metal-poor stars which we
expect to observe in different metallicity bins. The normalized distribution is
shown in Fig.~\ref{number} for bin sizes of 0.2 dex and 1.0 dex. The number
distribution is approximately a power law with slope 0.7 for [Fe/H] $< -3.5$ and
slope 0.9 for [Fe/H] $> -3.0$. Also plotted are the observed data from a
homogeneous intermediate resolution sample of Ryan \& Norris (1991) and the
combined high resolution data from Table~\ref{obs}, both rebinned to the large bin
size (see Tables~\ref{starnr} and \ref{starnr2} for the numerical values). It is
possible that the high resolution sample is incomplete in the range $[-3.0,-2.0]$,
whereas the intermediate resolution sample might have a shortage of stars with
[Fe/H] $< -3.0$. Table~\ref{starnr} also lists the number of model stars expected
in the three 1.0 dex bins, normalized such that the number of model stars in the
range $-4.0 <$ [Fe/H] $< -3.0$ is equal to the number of observed halo stars in
the high-resolution sample in this metallicity range. As can be seen, in this case
we expect $5\pm 2$ model stars with $-5.0 <$ [Fe/H] $< -4.0$ while the high
resolution sample contains none. If the ratio of stars in these two metallicity
bins for this admittedly inhomogeneous sample is representative for the Galactic
halo stars, this would suggest a genuine shortage of the most metal-poor stars. In
this case a possible solution could be the pre-enrichment of the halo ISM by
population III stars. It is conceivable that these already produced an iron
abundance of [Fe/H] $\approx -4.0$ before they disappeared, leaving only a
pre-enriched ISM.


\section{Conclusions \& Discussion}
\label{conclusions}

We have developed a stochastic model of the early chemical enrichment of the halo
ISM. The aim of the model is to understand the scatter in the [El/Fe] ratios of
observed stars at very low metallicities and the transition to the smaller scatter
seen at higher metallicities. We computed the evolution of the abundances of the
$\alpha$-elements O, Mg, Si and Ca, the iron-peak elements Cr, Mn, Fe and Ni and
the r-process element Eu and investigated the mixing of the halo ISM.

We divide the enrichment history of the halo ISM into different phases. At
metallicities [Fe/H] $< -3.0$, the ISM is not mixed and is dominated by local
abundance inhomogeneities, which are caused by individual Type~II SNe of different
progenitor masses. The second phase at [Fe/H] $> -2.0$ is defined by a well mixed
ISM which shows IMF averaged abundances and is too metal-rich to be dominated by
single SN events. A continuous transition from the first to the second phase
occurs between $-3.0 <$ [Fe/H] $< -2.0$. The onset of SN~Ia events marks the
beginning of a third phase in the enrichment of the ISM.

The different enrichment phases of the halo ISM can be distinguished in every
considered element, after normalising to the stellar yields. On the other hand,
Fig.~\ref{scatter} shows that some stellar yields reproduce the IMF averaged
element-to-iron ratio [El/Fe] well, but fail to reproduce the abundance scatter
observed in metal-poor halo stars. Especially the stellar yields of the
$\alpha$-elements O and Mg predict stars with a low element-to-iron ratio ([O/Fe]
or [Mg/Fe] $< 0.0$), which are not observed. Typically, metal-poor halo stars show
an overabundance of $\alpha$-elements of about [$\alpha$/Fe] $\approx 0.3$ to
$0.4$, as can be seen in Fig.~\ref{scatter}. This is especially troublesome, since
an attempt to solve this problem would either require a change in the iron yields
of the 13 and 15 M$_{\sun}$ models of up to a factor six, which would raise every
other mean element abundance by about 0.3 dex (a factor of two), or a change in
the stellar yields of oxygen and magnesium, which are produced mainly during the
hydrostatic burning phases.

Recent observations of metal-poor stars show a decrease of the [Cr/Fe] and [Mn/Fe]
ratios for lower metallicities. It is not possible to explain these trends with
the metallicity independent stellar yields we have used. Also, the stellar yields
of the iron-peak element Ni predict a scatter in [Ni/Fe] which is much too small
compared to the observational data. These problems argue strongly for a revision
of the theoretical nucleosynthesis models and their extension to lower
metallicities.

The unfortunate situation is, however, that there exists no theoretical foundation
to do so for Fe-group yields as long as the supernova explosion mechanism is not
understood. Thielemann et al. (\cite{th96}) discussed in detail uncertainties of
Fe-group yields due to the choice of mass cut, explosion energy and entropy, as
well as the delay time between collapse and explosion, affecting also the
neutron-richness of matter. Multidimensional aspects might add further degrees of
freedom (Nagataki et al. \cite{ng97}, \cite{ng98}). Thus, there is an
understanding of dependences, coincidences of abundance features etc., but at
present only observational information combined with galactic evolution modelling
like in the present paper or, e.g., Tsujimoto \& Shigeyama (\cite{ts98}) and
Nakamura et al. (\cite{na99}) can try to provide sufficient constraints for a
further understanding of supernova nucleosynthesis.

The advancing enrichment process of the halo ISM can be characterized by the
pollution factor $f_{\mathrm{poll}}$, defined as the ratio of the mass
$M_{\mathrm{poll}} = N_{\mathrm{SN}} \cdot M_{\mathrm{sw}}$ polluted by
$N_{\mathrm{SN}}$ preceding SNe and the ISM mass $M_{\mathrm{tot}}$, all in a unit
volume. The enrichment history of the halo ISM now is mainly determined by the
mixing efficiency which in turn is fixed by the ratio of the mass
$M_{\mathrm{sw}}$ of swept-up material in a SN event and $M_{\mathrm{tot}}$. The
more mass a SN sweeps up, the less SN events are needed to reach a certain value
of $f_{\mathrm{poll}}$, making the mixing more efficient. $M_{\mathrm{sw}}$ also
determines the average metallicity [Fe/H] of the halo ISM for a given pollution
factor. A larger swept-up mass leads to a lower mean ISM metallicity and vice
versa. Therefore, the metallicity where the transition from one enrichment stage
to the next occurs, depends only on the mixing efficiency
$M_{\mathrm{sw}}/M_{\mathrm{tot}}$ and not on the SFR. The SFR is only important
if one is interested in the \emph{elapsed time} that is needed to reach a certain
mean metallicity of the halo ISM.

Figures~\ref{sihist} and \ref{sidis} support our adopted value of $5 \times 10^4
\, \mathrm{M}_{\sun}$ of swept up material. If the swept up mass is higher, the
mixing would be more efficient, resulting in an IMF averaged chemical abundance
pattern at lower metallicities. This would produce a narrower peak in
Fig.~\ref{sihist} and a steeper slope in Fig.~\ref{sidis}. Moreover, the whole
curve in Fig.~\ref{sidis} would be shifted towards lower metallicities. On the
other hand, a smaller mixing mass would reduce the efficiency of the enrichment of
the ISM, which could be seen in a broader distribution at higher metallicities in
Fig.~\ref{sihist} and a shallower slope in Fig.~\ref{sidis}.

Standard 1-zone chemical evolution models predict a well defined age--metallicity
relation, based on the assumption that the ISM is well mixed at all times. In our
stochastic model the chemical inhomogeneity of the halo ISM and therefore the
scatter in metallicity at any time is much too high to reasonably establish such a
relation. Nevertheless, the steep rise with large scatter seen in
Fig.~\ref{agemetal} at very early times marks the chemically inhomogeneous
enrichment phase, while the slow increase later-on reflects the well mixed,
metal-rich ISM.

The results of our model also quantify the problem of the missing ultra metal-poor
stars. From it we have deduced the expected number of ultra metal-poor stars (with
[Fe/H] $< -4.0$) which should have been observed, normalized to the number of halo
stars in the combined high-resolution sample with metallicities in the range $-4.0
<$ [Fe/H] $< -3.0$. We expect about $5\pm2$ ultra metal-poor stars whereas none
was found to date. It is possible that Population III stars have caused a
pre-enrichment of the ISM to [Fe/H] $\approx -4.0$ and already have disappeared
before the onset of the formation of the Galaxy.

We would like to thank the referee S.~G. Ryan for helpful and motivating
discussions and for providing us with some of the observational data. This work
was supported by the Swiss Nationalfonds.



\begin{thebibliography}{}

\bibitem[1996]{ba96}
Balachandran, S.~C., Carney, B.~W., 1996, AJ, 111, 946

\bibitem[1991]{be91}
Bessel, M.~S., Sutherland R.~S., Ruan K., 1991, ApJ, 383, L71

\bibitem[1994]{be94}
Beveridge, R.~C., Sneden, C., 1994, AJ, 108, 285

\bibitem[1987]{bt87}
Binney, J., Tremaine, S., 1987, Galactic Dynamics, Princeton University Press,
Princeton, New Jersey

\bibitem[1999]{bo99}
Boesgaard, A.~M., King, J.~R., Deliyannis, C.~P., Vogt, S.~S., 1999, AJ, 117, 492

\bibitem[1991]{bo91}
Bouchet, P., Danziger, I.~J., Lucy, L.~B., 1991, AJ, 102, 1135

\bibitem[1996]{bu96}
Buchmann, L., 1996, ApJ, 468, L127

\bibitem[1997]{bu97}
Buchmann, L., 1997, ApJ, 479, L153

\bibitem[1990]{bu90}
Burrows, A., 1990, Ann. Rev. Nucl. Part. Sci., 40, 181

\bibitem[1995]{bu95}
Burrows, A., Hayes, J., Fryxell, B., 1995, ApJ, 450, 830

\bibitem[1996]{burr96}
Burrows, A., 1996, Nucl. Phys., A606, 151

\bibitem[1998]{bu98}
Burrows, A., Sawyer R.~F., 1998, Phys. Rev., C58, 554

\bibitem[1993]{cb93}
Charbonnel, C., Meynet, G., Maeder, A., Schaller, G., Schaerer, D., 1993, A\&AS,
101, 415

\bibitem[1996]{ca96}
Carney, B.~W., Laird, J.~B., Latham, D.~W., Aguilar, L.~A., 1996, AJ, 112, 668

\bibitem[1999]{ch99}
Chiappini, C., Matteucci, F., Beers, T.~C., Nomoto, K., 1999, ApJ, 515, 226

\bibitem[1998]{ch98}
Chiba, M., Yoshii, Y., 1998, AJ, 115, 168

\bibitem[1998]{chie98}
Chieffi, A., Limongi, M., Straniero, O., 1998, ApJ, 502, 737

\bibitem[1988]{ci88}
Cioffi, D.~F., McKee, C.~F., Bertschinger, E., 1988, ApJ, 334 252

\bibitem[1999]{co99}
Cowan, J.~J., Sneden, C., Ivans, I., Burles, S., Beers, T.~C., Fuller, G., 1999,
BAAS, 194, 67.04

\bibitem[1993]{fr93}
Fran\c{c}ois, P., Spite, M., Spite, F., 1993, A\&A, 274, 821

\bibitem[1999a]{fr99a}
Freiburghaus, C., Rembges, F., Rauscher, T., Kolbe, E., Thielemann, F.-K.,
Kratz, K.-L., Pfeiffer, B., Cowan, C., 1999, ApJ, 516, 381

\bibitem[1999b]{fr99b}
Freiburghaus, C., Rosswog, S., Thielemann, F.-K., 1999, ApJ 525, L121

\bibitem[1995]{fu95}
Fuhrmann, K., Axer, M., Gehren, T., 1995, A\&A, 301, 492

\bibitem[1989]{gr89}
Gratton, R.~G., 1989, A\&A, 208, 171

\bibitem[1991a]{gr91a}
Gratton, R.~G., Sneden, C., 1991a A\&A, 241, 501

\bibitem[1991b]{gr91b}
Gratton, R.~G., Sneden, C., 1991b A\&A, 287, 927

\bibitem[1997]{ha97}
Harris, W.~E., Bell, R.~A., Vandenberg, D.~A., Bolte, M., Stetson, P.~B.,
Hesser, J.~E., Van Den Bergh, S., Bond, H.~E., Fahlman, G.~G., Richer, H.~B., 
1997, AJ, 114, 1030

\bibitem[1999]{he99}
Heger, A., Langer, N., Woosley, S.~E., 1999, ApJ, astro-ph/9904132 (in press)

\bibitem[1984]{he84}
Henry, R.~B.~C., 1984, ApJ, 281, 644

\bibitem[1994]{he94}
Herant, M., Benz, W., Hix, W.~R., Fryer, C.~L., Colgate, S.~A., 1994, ApJ,
435, 339

\bibitem[1999]{ho99}
Hoffman, R.~D., Woosley, S.~E., Weaver, T.~A., Rauscher, T., Thielemann, F.-K.,
1999, ApJ, 521, 735

\bibitem[1999]{is99}
Ishimaru, Y., Wanajo, S., 1999, ApJ, 511, L33

\bibitem[1998]{is98}
Israelian, G., Garcia, L., Ram\'{o}n, J., Rebolo, R., 1998, ApJ, 507, 805

\bibitem[1994]{iw94}
Iwamoto, K., Nomoto, K., H\"oflich, P., Yamaoka, H., Kumagai, S., Shigeyama, T.,
1994, ApJ 437, L115

\bibitem[1998]{iw98}
Iwamoto, K., Mazzali, P.~A., Nomoto, K., Umeda, H., Nakamura, T., Patat, F.,
Danziger, I.~J., Young, T.~R., Suzuki, T., Shigeyama, T., Augusteijn, T.,
Doublier, V., Gonzalez, J.-F., Boehnhardt, H., Brewer, J., Hainaut, O.~R.,
Lidman, C., Leibundgut, B., Cappellaro, E., Turatto, M., Galama, T.~J.,
Vreeswijk, P.~M., Kouveliotou, C., Van Paradijs, J., Pian, E., Palazzi, E.,
Frontera, F., 1998, Nature 395, 672

\bibitem[1999a]{iw99a}
Iwamoto, K., 1999, ApJ, 512, L67

\bibitem[1999b]{iw99b}
Iwamoto, K., 1999, ApJ, 517, L47

\bibitem[1995]{ja95}
Janka, H.-T., M\"uller, E., 1995, Phys. Rep., 256, 135

\bibitem[1996]{ja96}
Janka, H.-T., M\"uller, E., 1996, A\&A, 306, 167

\bibitem[1999]{je99}
Jehin, E., Magain, P., Neuforge, C., Noels, A., Parmentier, G., Thoul, A.~A.,
1999, A\&A, 341, 241

\bibitem[1995]{ke95}
Keil, W., Janka, H.-T., 1995, A\&A, 296, 145

\bibitem[1998]{ko98}
Kozma, C., Fransson, C., 1998, ApJ, 497, 431

\bibitem[1994]{ki94}
King, J.~R., 1994, ApJ, 436, 331

\bibitem[1995]{la95}
Langer, N., Henkel, C., 1995. In: Busso, M., Gallino, R., Raiteri, C.~M. (eds.)
Nuclei in the Cosmos III. AIP Press, p.413

\bibitem[1997]{la97}
Langer, N., Fliegner, J., Heger, A., Woosley, S.~E., 1997, Nucl. Phys.,
A621, 457c

\bibitem[1988]{la88}
Larson, R.~B., 1988. In: R.~E. Pudritz, M. Fich (eds.) Galactic and Extragalactic
Star Formation. NATO ASI Series 232, 5, Kluwer, Dordrecht

\bibitem[1989]{ma89}
Magain, P., 1989, A\&A, 209, 211

\bibitem[1992]{ma92}
Mathews, G.~J., Bazan, G., Cowan, J.~J., 1992, ApJ, 391, 719

\bibitem[1999]{ma99}
Matteucci, F., Romano, D., Molaro, P., 1999, A\&A, 341, 458

\bibitem[1995]{mw95}
McWilliam, A., Preston, G.~W., Sneden, C., Searle, L., 1995, AJ, 109, 2757

\bibitem[1997]{mw97}
McWilliam, A., 1997, ARA\&A, 35, 503

\bibitem[1998]{mess98}
Messer, O.~E.~B., Mezzacappa, A., Bruenn, S.~W., Guidry, M.~W., 1998, ApJ,
507, 353

\bibitem[1997]{me97}
Meynet, G., Maeder, A., 1997, A\&A, 321, 465

\bibitem[1998]{mezz98}
Mezzacappa, A., Calder, A.~C., Bruenn, S.~W., Blondin, N.~J.~M., Guidry, M.~W.,
Strayer, M.~R., Umar, A.~S., 1998, ApJ, 495, 911

\bibitem[1990]{mb90}
Molaro, P., Bonifacio, P., 1990, A\&A, 236, L5

\bibitem[1990]{mc90}
Molaro, P., Castelli, F., 1990, A\&A, 228, 426

\bibitem[1999]{na99}
Nakamura, T., Umeda, H., Nomoto, K., Thielemann, F.-K.,, Burrows, A., 1999,
ApJ, 517, 193

\bibitem[1997]{ng97}
Nagataki, S., Hashimoto, M., Sato, K., Yamada, S., 1997, ApJ, 486, 1026

\bibitem[1998]{ng98}
Nagataki, S., Hashimoto, M., Sato, K., Yamada, S., Mochizuki, Y.~S., 1998, ApJ,
492, L45

\bibitem[1994]{ni94}
Nissen, P.~E., Gustafsson, B., Edvardsson, B., Gilmore, G., 1994, A\&A, 285, 440

\bibitem[1988]{no88}
Nomoto, K., Hashimoto, M., 1988, Phys. Rep., 163, 

\bibitem[1997]{no97}
Nomoto, K., Hashimoto, M., Tsujimoto, T., Thielemann, F.-K., Kishimoto, N.,
Kubo, Y., Nakasato, N., 1997, Nucl. Phys., A161, 79c13

\bibitem[1993]{no93}
Norris, J.~E., Peterson, R.~C., Beers, T.~C., 1993, ApJ, 415, 797

\bibitem[1990]{pe90}
Peterson, R.~C., Kurucz, R.~L., Carney, B.~W., 1990, ApJ, 350, 173

\bibitem[1999]{po99}
Pons, J.~A., Reddy, S., Prakash, M., Lattimer, J.~M., Miralles, J.~A., 1999,
ApJ, 513, 780

\bibitem[1994]{pr94}
Primas, F., Molaro, P., Castelli, F., 1994, A\&A, 290, 885

\bibitem[1997]{re97}
Reddy, S., Prakash, M., 1997, ApJ, 423, 689

\bibitem[1991]{ry91}
Ryan, S.~G., Norris, J.~E., Bessell, M.~S., 1991, AJ, 102, 303 

\bibitem[1991]{rn91}
Ryan, S.~G., Norris, J.~E., 1991, AJ, 101, 1865

\bibitem[1996]{ry96}
Ryan, S.~G., Norris, J.~E., Beers, T.~C., 1996, ApJ, 471, 254

\bibitem[1998]{sa98}
Saio, H., Nomoto, K., ApJ, 500, 388

\bibitem[1997]{sa97}
Samland, M., 1997, ApJ, 496, 155

\bibitem[1993a]{sr93a}
Schaerer, D., Meynet, G., Maeder, A., Schaller, G., 1993, A\&AS, 98, 523

\bibitem[1993b]{sr93b}
Schaerer, D., Charbonnel, C., Meynet, G., Maeder, A., Schaller, G., 1993,
A\&AS, 102, 339

\bibitem[1992]{sl92}
Schaller, G., Schaerer, D., Meynet, G., Maeder, A., 1992, A\&AS, 96, 269

\bibitem[1990]{sh90}
Shigeyama, T., Nomoto, K., 1990, ApJ, 360, 242

\bibitem[1998]{sh98}
Shigeyama, T., Tsujimoto, T., 1998, ApJ, 507, L 135

\bibitem[1994]{sn94}
Sneden, C., Preston, G.~W., McWilliam, A., Searle, L., 1994, ApJ, 431, L27

\bibitem[1998]{so98}
Sollerman, J., Leibundgut, B., Spyromilio, J., 1998, A\&A, 337, 207

\bibitem[1992]{sp92}
Spruit, H. C., 1992, A\&A, 253, 131

\bibitem[1991]{sw91}
Spiesman, W.~J., Wallerstein, G., 1991, AJ, 102, 1790

\bibitem[1991]{sp91}
Spite, M., Spite, F., 1991, A\&A, 252, 689

\bibitem[1992]{su92}
Suntzeff, N.~B., Phillips, M.~M., Elias, J.~H., Walker, A.~R., Depoy, D.~L.,
1992, ApJ, 384, L33

\bibitem[1994]{ta94}
Takahashi, K., Witti, J., Janka, H.-T., 1994, A\&A, 286, 857

\bibitem[1997]{ta97}
Talon, S., Zahn, J.-P., Maeder, A., Meynet, G., 1997, A\&A, 322, 209

\bibitem[1996]{th96}
Thielemann, F.-K., Nomoto, K., Hashimoto, M., 1996, ApJ, 460, 408

\bibitem[1998]{th98}
Thomas, D., Greggio, L., Bender, R., 1998, MNRAS, 296, 119

\bibitem[1999]{tr99}
Travaglio, C., Galli, D., Gallino, R., Busso, M., Ferrini, F, Staniero, O.,
1999, ApJ, 521, 691

\bibitem[1998]{ts98}
Tsujimoto, T., Shigeyama, T., 1998, ApJ, 508, L151

\bibitem[1998]{tu98}
Turatto, M., Mazzali, P.~A., Young, T.~R., Nomoto, K., Iwamoto, K., Benetti, S.,
Cappallaro, E., Danziger, I.~J., De Mello, D.~F., Phillips, M.~M.,
Suntzeff, N.~B., Clocchiatti, A., Piemonte, A., Leibundgut, B., Covarrubias, R.,
Maza, J., Sollerman, J., 1998, ApJ 498, L129

\bibitem[1999]{um99}
Umeda et al., 1999, in \emph{First Stars}, ESO (in press)

\bibitem[1996]{wa96}
Wasserburg, G., Busso, M., Gallino, R., 1996, ApJ, 466, L109

\bibitem[1998]{wh98}
Wheeler, J.~C., Cowan, J.~J., Hillebrandt, W., 1998, ApJ 493, L101

\bibitem[1994]{wo94}
Woosley, S.~E., Wilson, J.~R., Mathews, G.~J., Hoffman, R.~D., Meyer, B.~S.,
1994, ApJ, 433, 229

\bibitem[1995]{ww95}
Woosley, S.~E., Weaver, T.~A., 1995, ApJS, 101, 181

\bibitem[1999]{ya99}
Yamada, S., Janka, H.-T., Suzuki, H., 1999, A\&A, 344, 533

\bibitem[1990]{zh90}
Zhao, G., Magain, P., 1990, A\&A, 238, 242

\end{thebibliography}
\end{document}